# Impairing the memory of an electron-glass by IR excitation

V. Orlyanchik, A.Vaknin and Z. Ovadyahu
*The Racah Institute of Physics*
*The Hebrew University, Jerusalem 91904 Israel*

and M. Pollak
*Department of Physics, University of California, Riverside CA 92651, USA*

## Abstract

We study the influence of various excitations on the anomalous field effect observed in insulating indium-oxide films. In conductance G versus gate-voltage $V_g$ measurements one observes a characteristic cusp around the $V_g$ at which the system has equilibrated. In the absence of any disturbance this cusp may persist for a long time after a new gate voltage was imposed on the sample and hence reflects a memory of the previous equilibrium state. This memory is believed to be related to the correlations between electrons. Here we show that exciting the conduction electrons by exposing the sample to IR light degrades this memory. We argue that any excitation that randomizes the system destroys the correlations and therefore impairs the memory.

## Introduction

It has been predicted[1] that Anderson insulators with interactions may lead to a glassy state. Evidence for non-ergodic behavior in such systems has been recently found in several experiments.[2,3] These include slow evolution of the conductivity following an excitation and various memory effects.

An example of such a memory is revealed in field-effect (FE) experiments[3]. After quench-cooling the sample to low temperature with a certain gate-voltage $V_g^o$ a cusp slowly develops, in the $G(V_g)$ traces[4] centered at $V_g^o$. Following such an equilibration process, the system 'remembers' $V_g^o$ even after the gate-voltage was fixed at another value for some time $\tau$. This memory can be revealed in subsequent measurements of $G(V_g)$ as a cusp-like minimum centered at $V_g^o$. Without any further disturbance, the amplitude of this 'memory cusp' (MC) decreases logarithmically with the time $\tau$, and can be detected even after many hours.

In this paper we study the effects of various external excitations on the MC. We show that exposure of the sample to IR light considerably reduces the amplitude of the MC and gives rise to slow relaxation of conductivity. This slow relaxation might be viewed as a rebuilding process of this cusp. These findings are consistent with the conjecture that memory in these systems is related to the interactions between electrons. We also find that the response of the system to excitation by MW is quite different and discuss the possible implication of this result.

## Experimental

The samples used in this study were prepared in a MOSFET-like configuration. The active layer was a 50 Å thick polycrystalline $In_2O_{3-x}$ film e-gun deposited using 99.999% pure $In_2O_3$ onto a microscope cover-glass substrate (100-140 μm thick). Data is shown for two samples: Sample #1 has lateral dimensions of 2x2 mm, and sample #2 has lateral dimensions of 0.5 mm (width) and 1.6 mm (length). The resistivity of these samples could be varied by UV treatment.[5] The gate, a 500 Å thick gold film, was evaporated *vis-à-vis* the $In_2O_{3-x}$ film on the other side of the glass. Full details of the sample preparation and characterization of the $In_2O_{3-x}$ films used in this study are described elsewhere.[5]

Measurements were carried out at T=4.11 K with the samples immersed in liquid $^4$He storage dewar. Temperature changes were achieved by lifting the sample slightly above the $^4$He level and monitoring the increase in T with a Ge thermometer. The conductivity of the samples was measured using a two terminal ac technique employing an ITHACO 1211 current amplifier and a PAR 124A lock in amplifier or by using 100 kΩ resistor connected in series with the sample and a 5204 lock in amplifier. Care was taken to ensure linear-response by the use of sufficiently low ac bias. A LED device was mounted on the sample stage in front of the sample and used as a source of infrared (IR) radiation ($\upsilon=3 \cdot 10^{14}$ Hz). The LED output-power was measured, at room temperature, by a thermopile device (Newport 815). A Gunn-diode was used as a source of microwave (MW) radiation ($\upsilon=10^{11}$ Hz). In the experiments involving excitation by MW the sample was mounted at the bottom of a cylindrical (10 mm diameter) stainless-steel probe that formed a cavity for the microwave radiation.

## Results and discussion

The effect of excitation by IR light on the MC is shown in Figure 1a. In these experiments we first quench-cooled the sample to the measurement temperature $T_m$=4.1 K with a fixed gate voltage $V_g^o$=0 and allowed the system to equilibrate for about a day under these conditions. Then, at t≡0, while recording $G(V_g)$ we swept $V_g$ from $V_g^o$ to $V_g^n$=106 V, kept it there for a time $\tau$=15 seconds, and swept it back to $V_g$=−106 V. The resulting trace reveals a cusp in $G(V_g)$ centered at $V_g$=0, which is the MC alluded to in the introduction. Then, the same procedure was used except that during the first 5 seconds of the waiting period $\tau$ we exposed the sample to IR radiation. Evidently, the exposure to IR light affects the memory of the system; as the IR intensity increases, the amplitude of the MC is monotonically reduced and it essentially vanishes for the highest intensity used.

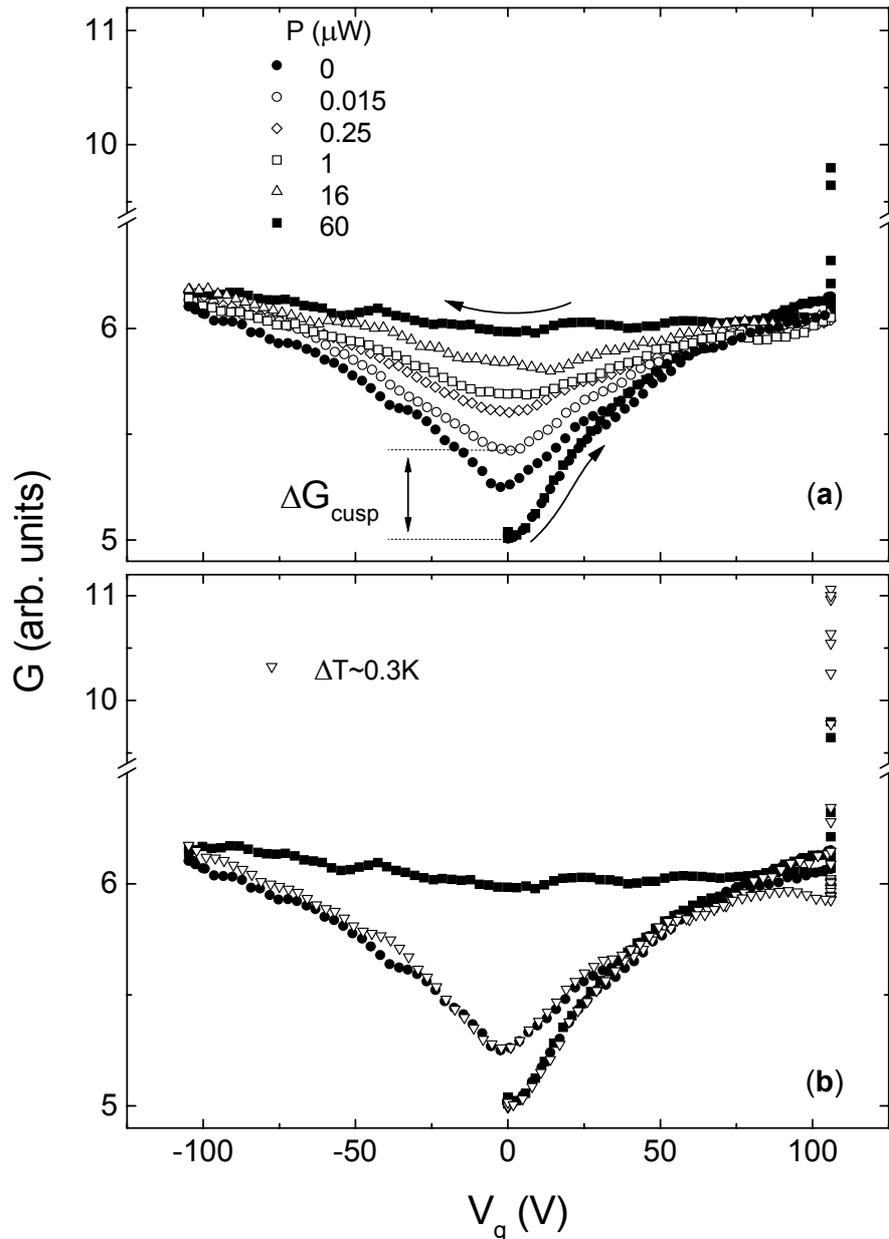

**Figure 1**. (**a**) Field effect traces measured after a brief exposure to IR radiation with various power levels P. Each trace starts by sweeping $V_g$ from zero to $V_g$ =106 V (as indicated by the lower arrow). After $\tau$=15 seconds (and IR exposure), $V_g$ is swept back to –106 V (as indicated by the upper arrow). Also shown is the definition of $\Delta G_{cusp}$ for the P=0.5 μW trace. (**b**) Field effect trace measured as above except that during the waiting time $\tau$ we raised the temperature for a few seconds then rapidly re-cooled to $T_m$. The resulting $G(V_g)$ trace is compared with two of the traces taken from (**a**). Sweep rate is 2 V/sec in all cases. Sample #1, $R_\square$=600 MΩ.

These results are not due to heating. To achieve an effective MC degradation it is *not* sufficient to introduce a disturbance that increases the system conductance by a certain amount. For example, raising the sample temperature for a few seconds by ΔT (chosen such that the peak increase in conductivity due to ΔT is comparable with the increase in conductivity due to the IR illumination), does not affect the MC. This is shown in figure 1b. Note that in this case the memory was not degraded, in contrast with the considerable degradation caused by the IR radiation.

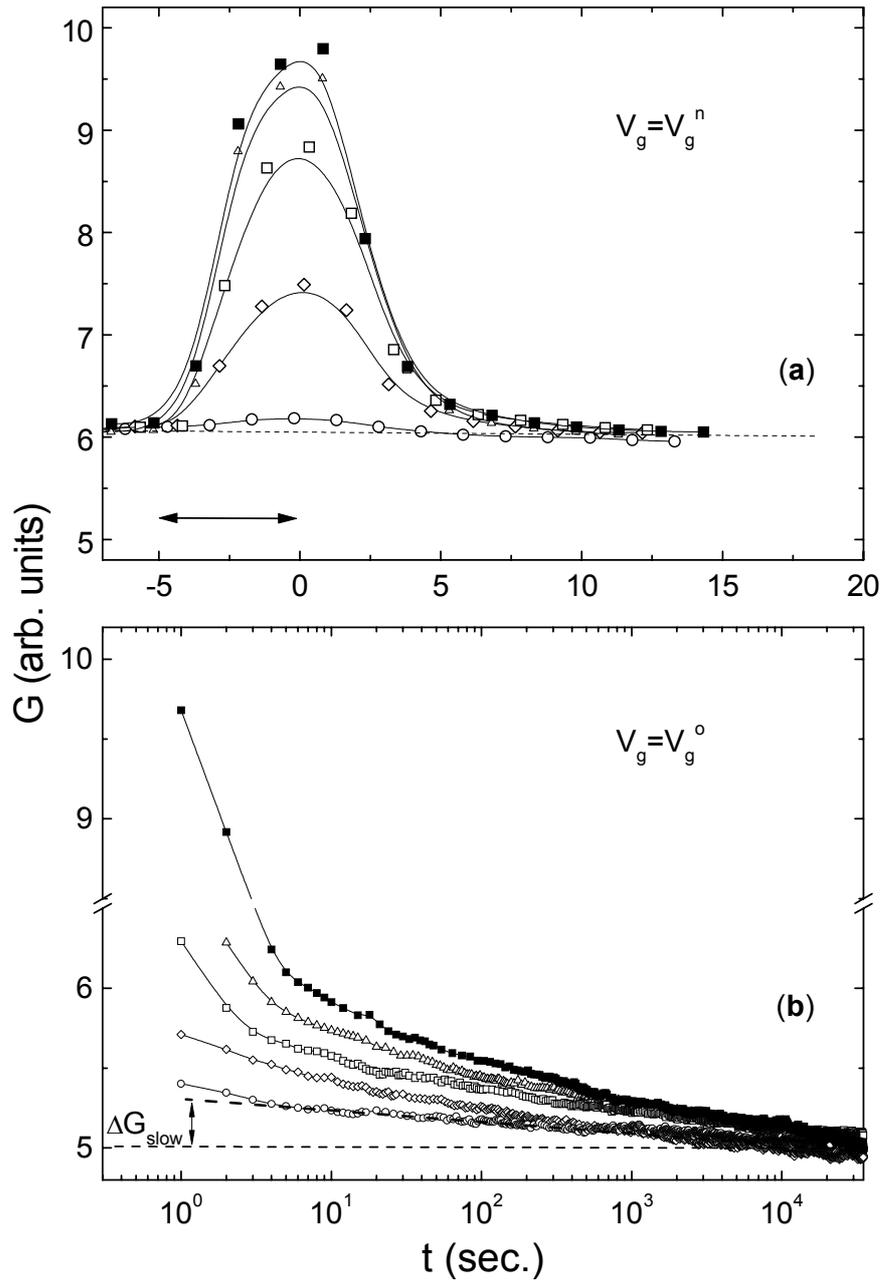

**Figure 2**. (**a**) The conductance versus time during (interval marked by the arrow) and after exposure to electromagnetic radiation with $\upsilon=3\cdot10^{14}$ Hz (IR). The various traces correspond to different power levels P, as specified in Figure 1. During the measurement $V_g$ was fixed at 106 V. (**b**) Conductance versus time after exposing the sample to IR radiation as in (**a**) except that during this measurement $V_g$ was fixed at 0 V. Dashed lines mark the value of G prior to the exposure. The sample is the same as in Figure 1.

Figure 2 shows the time dependence of the conductivity, following an exposure to IR radiation, for two experiments: In one experiment (figure 2b) the exposure occurs from $V_g^o$, where the system was equilibrated. In the other (figure 2a) the exposure occurs just after a sweep of $V_g$ from $V_g^o$ to $V_g^n$. As seen in the figure, there is a notable difference between these two cases. In figure 2a, the conductivity rapidly reaches the value it had prior to the excitation. In figure 2b, on the other hand, after an initial rapid decay the conductivity remains at a higher value than it had before the excitation. As seen in the figure, it takes a very long time for this excess conductivity to decay.

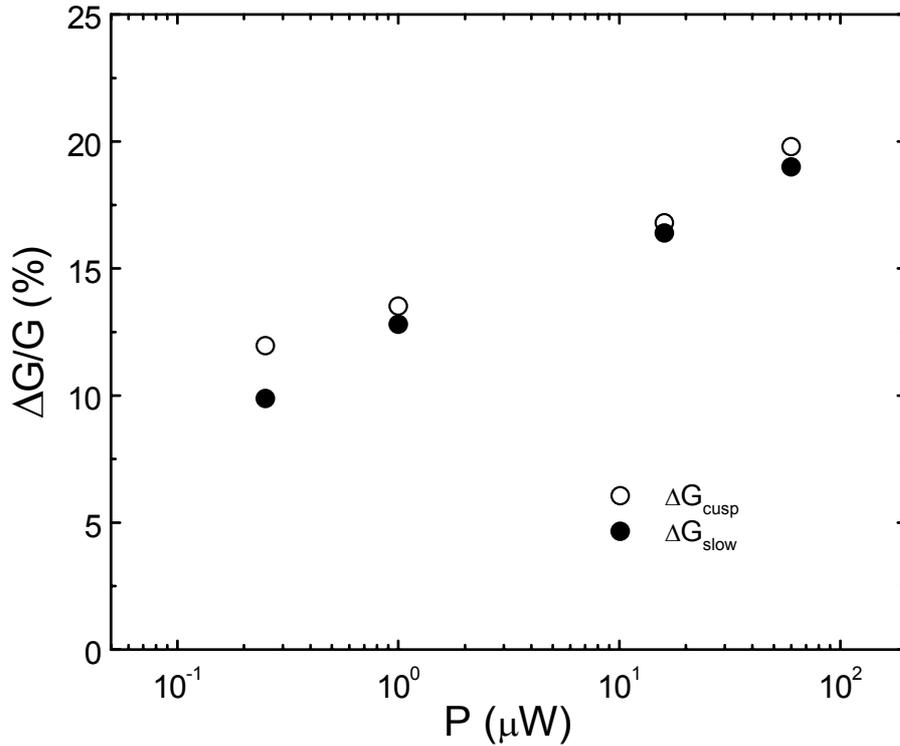

**Figure 3**. $\Delta G_{slow}$ and $\Delta G_{cusp}$ as function of the IR power. The definitions of these quantities are somewhat arbitrary in the sense that $\Delta G_{slow}$ is defined at t=1 second, and $\Delta G_{cusp}$ depends on the sweep-rate (w). However, owing to the logarithmic dependences of $\Delta G_{slow}(t)$ and $\Delta G_{cusp}(w)$ the correlation between $\Delta G_{slow}$ and $\Delta G_{cusp}$ is not sensitive to the particular value of time and of sweep-rate we use.

We now show that the slow component of this excess conductance $\Delta G_{slow}$ correlates with the change in the cusp amplitude $\Delta G_{cusp}$ due to the IR excitation. These quantities are defined as follows. $\Delta G_{slow}$ is the amount by which the conductivity measured 1 second after the excitation has been removed exceeds its equilibrium value. This was obtained by extrapolation of the slowly relaxing part of the conductivity (c.f., figure 2). $\Delta G_{cusp}$ is the difference between the conductivity at the minimum of the MC (at $V_g^o$) and the initial conductivity at t=0 and $V_g=V_g^o$ (see figure 1). These quantities are shown in figure 3 for the various intensity levels used in this experiment.

As explained below, $\Delta G_{slow}$ and the re-building process of the memory cusp of amplitude $\Delta G_{cusp}$ are two aspects of the same underlying physics. (We note that the measured value of $\Delta G_{cusp}$ is not affected by the fast relaxation because it is measured long enough after irradiation.) As figure 3 clearly shows, the values of $\Delta G_{slow}$ and $\Delta G_{cusp}$ in two different types of experiments are similar. The difference between them for small P is essentially due to the natural degradation of $\Delta G_{cusp}$ during $\tau$. It is also important to note that no $\Delta G_{slow}$ can be produced unless prior to the excitation the system was allowed to equilibrate and build a cusp at the particular value of $V_g$ used in the measurement (figure 2).

Next, we compare the effect of exposing the system to MW and IR radiations (Figure 4a). The figure shows the conductivity as a function of time before, during, and after the excitations. In both experiments the system was allowed to equilibrate for 24 hours prior to the excitation. Although these perturbations were kept for the same time and their intensity was adjusted to get the same $\Delta G$, the ensuing relaxation behavior is quite different. Following the IR exposure there is a considerable component of $\Delta G_{slow}$ while after the exposure to MW radiation the conductivity rapidly returns to its equilibrium value. In addition, non-stationary response of the conductivity is apparent also *during* the excitation in the case of IR radiation but not during the MW excitation.

The question is what is the reason for the difference between exposing the system to IR and MW radiation. Obviously, the only difference between the two types of electromagnetic fields is their photon energies. This leads us to conjecture that to obtain an excited state from which the system relaxes in a sluggish manner, the quantum of energy associated with the perturbation must exceed a certain value. The latter is presumably associated with the correlation energy $E_c$ responsible for the glassy state. It has been suggested before that the width of the cusp $\Gamma$ observed in the FE experiments may be a measure of this energy. For our systems, $\Gamma$ is of the order of 10-40 K. This should be compared with the energy $h\upsilon$ for the MW and IR which are 5 K and $1.5 \cdot 10^4$ K respectively. This is consistent with our conjecture but it is clearly desirable to get a better handle on the threshold energy needed to excite the system.

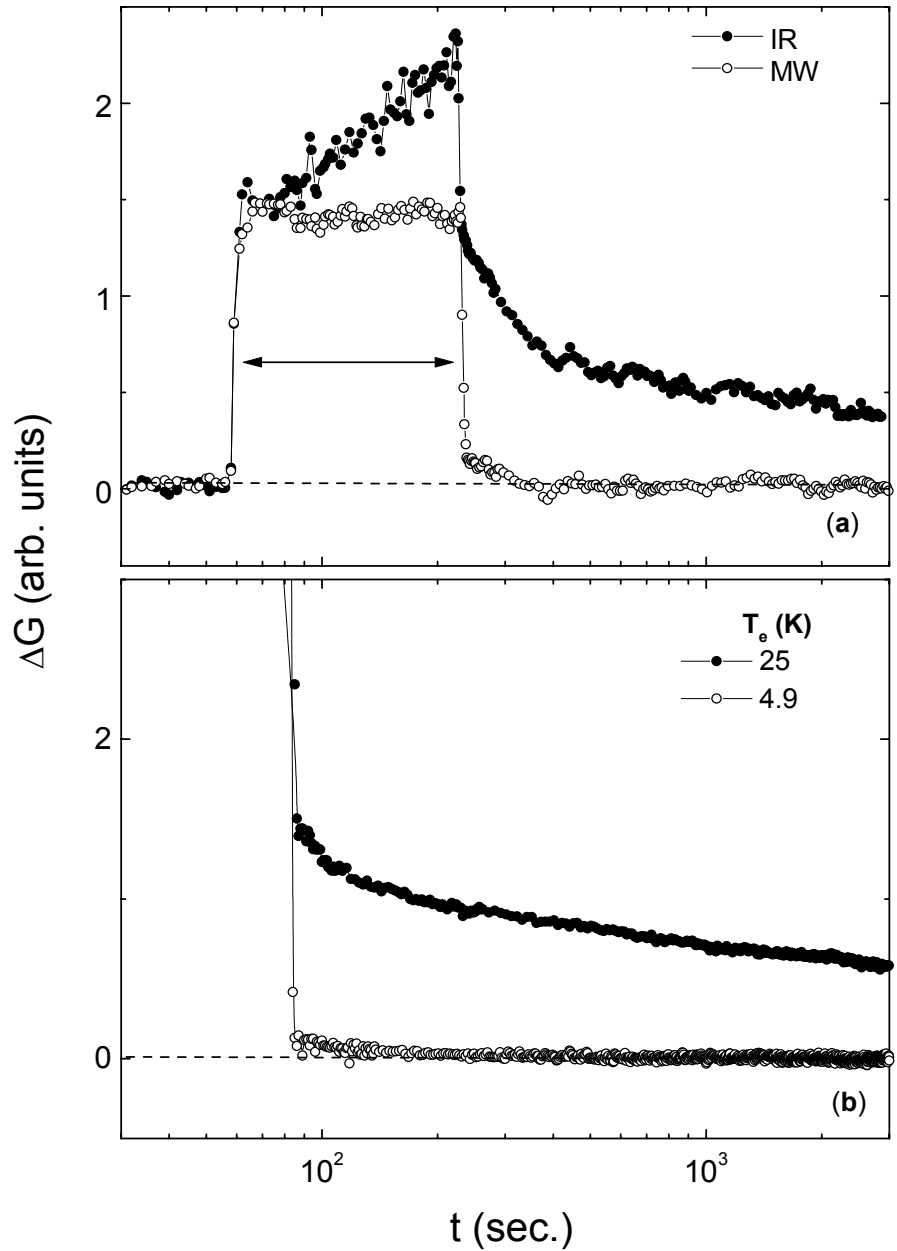

**Figure 4**. (**a**) The excess conductance versus time during (interval marked by the arrow) and after exposure to electromagnetic radiation with $\upsilon$=100 GHz (MW) and $\upsilon$=3·10$^{14}$ Hz (IR). Sample #2, $R_\square$=60 M$\Omega$ (**b**) The excess conductance versus time following brief heating of the sample to $T_e$. Sample #2, $R_\square$=320 M$\Omega$.

To narrow down this range of energies, we performed a series of experiments in which the sample temperature was briefly raised to $T_e>T_m$, and $\Delta G(t)$ was monitored after re-cooling to $T_m$ for various values of $T_e$. Figure 4b shows an example for such a procedure. It is seen that for $T_e \sim 25K$, $\Delta G(t)$ shows a substantial component of $\Delta G_{slow}$ which is similar to the case of IR exposure. For $T_e \sim 5K$, on the other hand, $\Delta G_{slow}$ is barely observed above the noise level. Note that, due to its statistical nature, imparting energy to the system by raising the temperature may provide energy-quanta that are somewhat larger than $T_e$. This is in contrast to the IR and MW experiments where there is a well-defined quantum of energy. Nonetheless, it seems fair to say that the experiments in figure 4b demonstrate that energy of the order of 10-25 K is necessary to drive the system far from equilibrium. This, in turn, is consistent with the assumption that $E_c$ is of the order of the cusp width $\Gamma$. These findings are also in line with the temperature dependence of the cusp in $G(V_g)$. The amplitude of the cusp is quickly diminished with T, and typically vanishes[6] above 10-40 K.

In summary, we have shown that exciting the system may impair the memory effects in the electron glass. We compared the effects produced by various mechanisms (IR and MW radiations, and an increase of the bath temperature), on the glassy transport properties of the system. This led us to conclude that there is a characteristic energy underlying the non-ergodic effects observed in the studied system.

In the context of the view[7] that the memory is related to the correlations between electrons, we argue that any excitation that results in a random occupation of electronic states should destroy these correlations and thus the memory. Our results suggest that there is a threshold energy necessary to achieve such an effect. It is then natural to associate this energy with the electron-electron correlation energy that must be overcome in order to produce randomness in the occupation of electrons.

This research was supported by a grant administered by the US-Israel Science Foundation and a grant administered by the German-Israel Science Foundation.